\documentclass[aps,prb,twocolumn,superscriptaddress, longbibliography]{revtex4-2} 
\usepackage{graphicx}  
\usepackage{bm}        
\usepackage{amssymb}   
\usepackage{amsmath}
\usepackage{hyperref}
\usepackage{natbib}

\usepackage{lineno}

\hypersetup{
	colorlinks=true,
	urlcolor = blue,
	citecolor=blue,
	linkcolor= blue,
	bookmarks=true,
	bookmarksopen=false,
}

\begin{document}

\title{Nonreciprocal Local-Resonance Induced Complex Band Hybridization}
\author{Wang Tat Yau} \affiliation{Department of Physics, The Hong Kong University of Science and Technology, Hong Kong, China}
\author{Kai Fung Lee} \affiliation{Department of Applied Physics, The Hong Kong Polytechnic University, Hong Kong, China}
\author{Raymond P. H. Wu} \affiliation{Department of Physics, The Chinese University of Hong Kong, Hong Kong, China}
\author{Wai Chun Wong} \affiliation{Department of Physics, The Hong Kong University of Science and Technology, Hong Kong, China}
\author{Jensen Li} \affiliation{Department of Physics, The Hong Kong University of Science and Technology, Hong Kong, China}
\author{C. T. Chan} \affiliation{Department of Physics, The Hong Kong University of Science and Technology, Hong Kong, China}
\affiliation{Institute for Advanced Study, The Hong Kong University of Science and Technology, Hong Kong, China}
\author{Kin Hung Fung} \email{funguiuc@gmail.com}\affiliation{Department of Physics, The Hong Kong University of Science and Technology, Hong Kong, China}
\date{\today}

\begin{abstract}
We study the complex band hybridization induced by nonreciprocal local resonances in photonic crystals. Composed of trimer unit cells, a two-dimensional (2D) magnetophotonic crystal with an analytically obtainable solution is considered. We find that nonreciprocal spectral gap may appear without nonreciprocal transmission and that the imaginary parts of the complex wavevectors $\text{Im}(\mathbf{k})$ may blow up at resonance to give extreme nonreciprocal transmission. We further show that, for a subwavelength lattice, the isolation ratio for the nonreciprocal transmission is determined solely by $\text{Im}(\mathbf{k})$ instead of the extensively studied real part $\text{Re}(\mathbf{k})$. Our finding contradicts the common belief that ``spectral nonreciprocity [$\omega(\mathbf{k})\neq\omega(-\mathbf{k})$] always implies nonreciprocal transmission".
\end{abstract}

\maketitle
Band hybridization in solid-state materials plays a significant role in the transport properties of waves and particles. Introducing additional localized states in crystals can further induce many interesting band properties and transport phenomena. For example, Kondo insulator effects~\cite{kondo1,kondo2} occur when electronic bandgaps are created through the hybridization between localized states and conduction bands. Similar band hybridization effects can also occur in artificial wave-functional materials, leading to novel wave transport phenomena such as local resonance gaps~\cite{pni1,phi4}, superabsorption~\cite{abs3,abs0}, extraordinary transmission~\cite{et0}, and slow waves~\cite{bband}. Recent literature has also reported that local resonance flat bands may have non-trivial topological nature~\cite{fbt1,fbt2}, which has played a unique role in robust (nonreciprocal) one-way transport against disorder~\cite{TP1,TP2}.

Nonreciprocal transport of light requires the breaking of electromagnetic reciprocity (i.e., nonreciprocity)~\cite{NR1,NR2}, which refers to the difference in the local electromagnetic fields received when the source and the receiver are exchanged. For photonic applications such as in isolators~\cite{L_NR}, circulators~\cite{cir0}, and directional amplifiers~\cite{da0}, electromagnetic nonreciprocity must be severely broken. Current mechanisms for nonreciprocity include nonlinearity~\cite{NL_NR}, time modulation~\cite{T_NR} and Lorentz nonreciprocity~\cite{L_NR}. In the linear regime, periodic photonic structures composed of Lorentz nonreciprocal media can support spectral nonreciprocity~[i.e., $\omega(\mathbf{k})\neq\omega(-\mathbf{k})$ for real values of angular frequency $\omega$ and wavevector $\mathbf{k}$] when all necessary symmetries are broken~\cite{spec_non, rot_temp, pt_muzero}.

\begin{figure} 
    \centering
        \begin{tabular}{@{}c@{}}
            \includegraphics[width=0.49\linewidth]{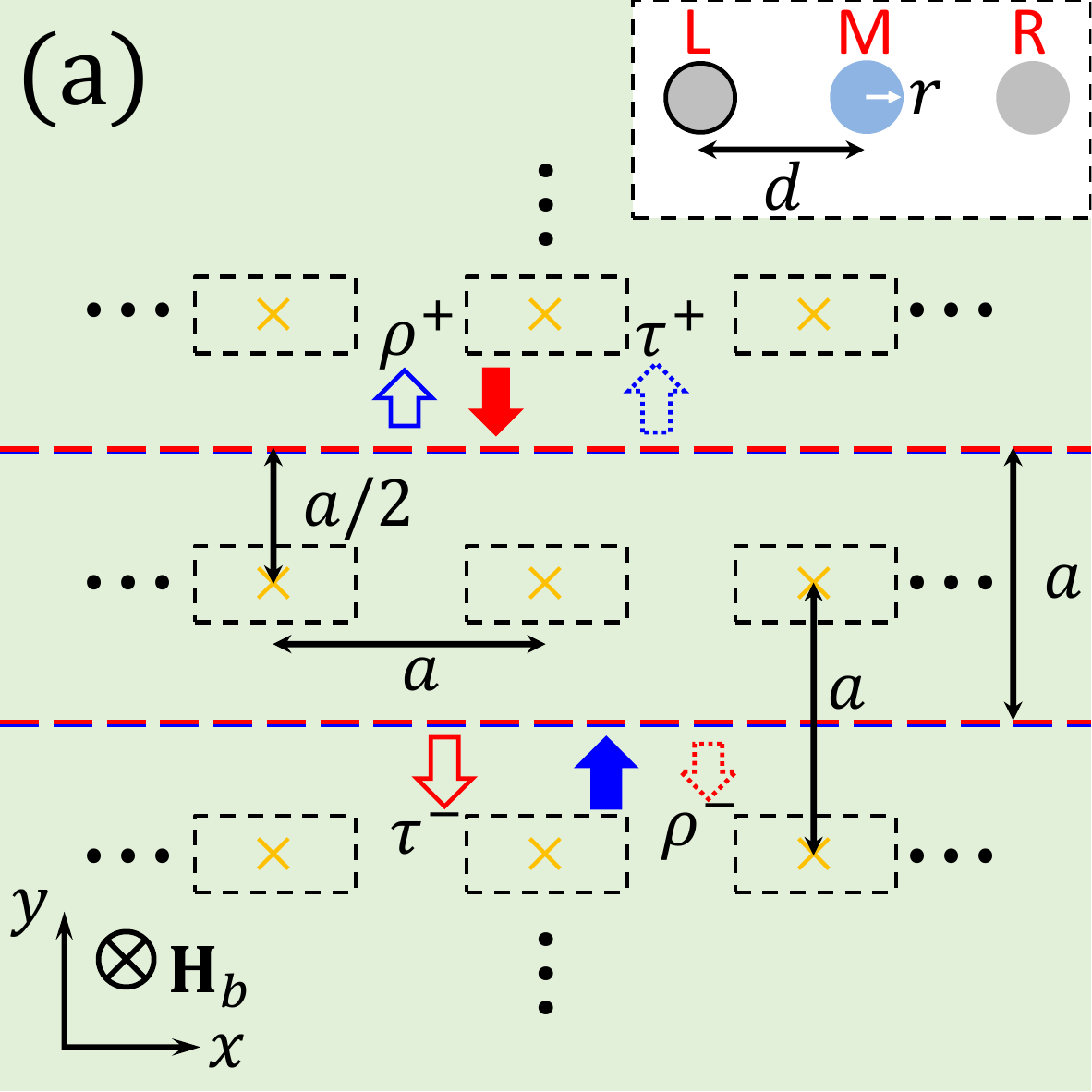}
        \end{tabular}
        \begin{tabular}{@{}c@{}}
            \includegraphics[width=0.49\linewidth]{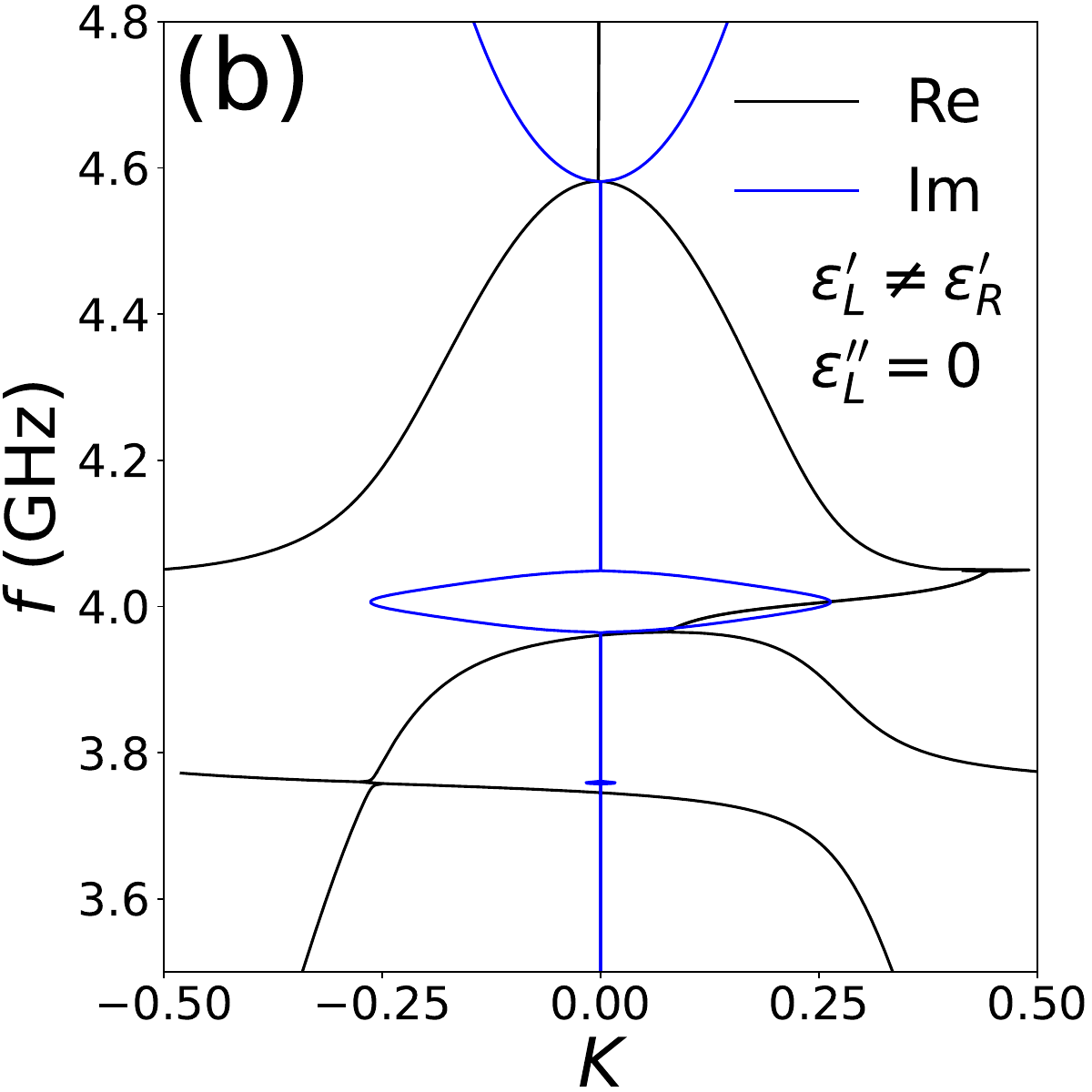}
        \end{tabular}
	\caption {\small (color online) Band hybridization due to nonreciprocal local resonance. (a) Lattice structure of the 2D trimer photonic crystal. The yellow crosses represent the centers of unit cells shown in the inset, which consists of a gyromagnetic cylinder~(blue circle) at the center and the two dielectric cylinders~(gray circle) on two sides. The hollow arrows with solid (dashed) boundaries are the incoming (outgoing) waves for each layer. The red (blue) solid arrows represent the forward (backward) incident directions.  (b) Complex band structure for the nontrivial but Hermitian case. $K$ is the normalized complex wavenumber in $y$-direction. The black and blue lines refer to the real and imaginary parts of $K$, respectively. Geometric parameters are lattice constant $a=50$ mm, core-to-core distance $d=12$ mm and radius $r=1~$mm. Material parameters are $M_{s}=1750~$Oe, $H_{0}=500~$Oe, $\varepsilon_M=15$, $\varepsilon_R'=92.16$, $\varepsilon_L'=\varepsilon_R'/1.44$, $\varepsilon_L''=\varepsilon_R''=0$.~\cite{note1,tms2,ceramic}}
    \label{scm}
\end{figure} 

In contrast, transport properties are especially sensitive to the complex nature of wavevectors~\cite{cbs,cbs_0,cbs_2}, which was largely ignored in Bloch's theory of Hermitian systems~\cite{nonBloch,nonBloch2,nonBloch3}. Similarly to the reciprocal cases, nonreciprocal local resonators could also provide peaks of $\text{Im}(\mathbf{k})$ in the real frequency spectra~\cite{nrpc}, and a nonreciprocal active metamaterial~\cite{nm} has experimentally shown a large difference between $\text{Im}(\mathbf{k})$ for nonreciprocal transmission. This may suggest that there is a fundamental relation between $\text{Im}(\mathbf{k})$ and nonreciprocal transmission that has not been explicitly shown. In this Letter, we study the realization of such nonreciprocal features on the complex band structure (CBS)~\cite{cbs,cbs_0,cbs_2} (with complex values of $\mathbf{k}$) and the associated physical consequences by considering a two-dimensional (2D) trimer photonic crystal made of gyromagnetic and dielectric cylinders [see Figure~\ref{scm}(a)]. 

We consider the band hybridization among nonreciprocal local resonances and band-folded propagating photon bands. The nonreciprocal local resonances originate from the gyromagnetic cylinder placed at the center in each unit cell [see the label M in Fig.~\ref{scm} (a)]. The two side-by-side parallel dielectric cylinders (L and R) are to provide tunable degrees of freedoms to turn on/off the spatio-temporal symmetries and Hermiticity. These cylinders couple with the gyromagnetic cylinders to provide hybridized modes in each single grating layer~\cite{tms2}. The relative permittivity of these cylinders is denoted by $\varepsilon = \varepsilon'+i\varepsilon''$, where $\varepsilon'$ and $\varepsilon''$ correspond to the real and imaginary parts, respectively. A subscript of M, L and R is used to label corresponding parameters for the gyromagnetic, left and right cylinders, respectively. The dielectric properties of the left cylinder are used to control the breaking of spatial symmetries and hermiticity through $\varepsilon_L'$ and $\varepsilon_L''$, respectively. The gyromagnetic cylinder is subjected to a bias magnetic field in the negative $z$ direction (along the cylinder axis) and its permeability tensor is used to control the breaking of Lorentz reciprocity. For continuous wave of single angular frequency $\omega=2\pi f$ (with $f$ being the real frequency), its relative permeability tensor is given as~\cite{phi1}
\begin{equation}
\label{mut}
	\boldsymbol{\mu}=\left(
		\begin{tabular}{ccc}
		$\mu_r$&$i\mu_\kappa$&$0$\\
		$-i\mu_\kappa$&$\mu_r$&$0$\\
		$0$&$0$&$1$\\
		\end{tabular}
	\right),
\end{equation}
where $\mu_r = 1+\omega_{m}\omega_{h}/(\omega_{h}^{2}-\omega^{2})$, $\mu_\kappa = \omega_{m}\omega/(\omega_{h}^{2}-\omega^{2})$, $\omega_{m} = \gamma M_{s}$, $\omega_{h}=\gamma H_{0}$, $\gamma$ is the gyromagnetic ratio, $M_{s}$ is the saturation magnetization in the ferromagnetic materials and $H_{0}$ is the applied static magnetic field. Here, the time harmonic convention is taken as $e^{-i\omega t}$. This permeability tensor describes a Lorentz nonreciprocal medium with $\boldsymbol{\mu}\neq\boldsymbol{\mu}^\text{T}$~\cite{jakong}. The parameter $\mu_\kappa$ could be later artificially set to zero for comparison with the Lorentz reciprocal cases. A small cylinder made of such material could provide nonreciprocal local resonance at the frequency where $\mu_r+\mu_\kappa\approx-1$~\cite{phi1}. When there is no loss/gain media (i.e. $\mu_r$, $\mu_\kappa$, and all $\varepsilon$ are real numbers), the permeability tensor satisfies $\boldsymbol{\mu}=\boldsymbol{\mu}^{\dagger}$ and the photonic (magnon) system is Hermitian~\cite{hermitmagnon}.

We consider transverse electric modes with electric fields parallel to the $z$-axis. The dynamic responses of the dielectric and gyromagnetic cylinders are, respectively, modeled as electric displacement currents oscillating along the $z$-axis and magnetization precessing about $z$-axis. We describe the 2D photonic crystals as layers of gratings stacked along the $y-$direction, as shown in Fig.~\ref{scm}(a). To obtain analytical solutions, multiple scattering theory~\cite{tms1,tms2} is applied to all cylinders within each layer and transfer/scattering matrix method~\cite{bs,Supp} is used to account for the multiple scatterings between layers. The grating constant $a$ in each layer is considered smaller than the background wavelength so that $\omega a/c <2\pi$ and all non-zeroth-order diffractions are evanescent. Neglecting evanescent coupling between layers~\cite{Comment1}, each transfer/scattering matrix becomes a simple $2\times2$ matrix and the (generalized) Bloch’s theorem gives the dispersion relations, which can be written as~\cite{Supp}
\begin{equation}\label{eqn:DRC}
\begin{cases} 
    \cos(k^\pm_ya) = \frac{T_r}{4}\left(1+{D_t}^{-1}\right)\pm\left(1-{D_t}^{-1}\right)\sqrt{T_r^2/4-D_t}\quad\\
    \sin(k^+_ya)+\sin(k^-_ya)=\frac{T_r}{2i}\left(1-D_t^{-1}\right),\\
\end{cases}
\end{equation}
where $k_y^\pm$ are the y-components of the two different solutions of the wavenvectors. We solve for complex number solutions of $k_y^\pm$ satisfying Eq.~(\ref{eqn:DRC}) by keeping the frequency $\omega$ a real number. The choice of solutions with real frequency and complex wavevectors has an advantage of modeling continuous wave excitation, which was employed in some reciprocal systems~\cite{cbs,cbs_0,cbs_2}. In Eq.~(\ref{eqn:DRC}), $T_r=$ tr(\textbf{C}) and $D_t=$ det(\textbf{C}) are, respectively, the trace and the determinant of the frequency-dependent transfer matrix~\cite{Supp}
\begin{equation}
	\mathbf{C}=\frac{1}{\tau^-}
		\left[
			\begin{tabular}{cc}
				$(\tau^+\tau^--\rho^+\rho^-)e^{ik_0a}$&$\rho^+$\\
				$-\rho^-$&$e^{-ik_0a}$
			\end{tabular}
	\right],\label{eqn:Cd}
\end{equation}
for a single layer of cylinder grating lying on the $xz$ plane, and $k_0=\omega/c$ refers to the free-space wavenumber. We note that $\rho^+=\rho^-=\rho$ due to an unobvious spatiotemporal symmetry of the grating~\cite{Supp}. The determinant $D_t=\tau^+/\tau^-$ indicates transmission nonreciprocity of single grating layer. When $D_t=1$, the closed-form expression reduces back to the reciprocal cases as in the literature~\cite{rec_eff,rec_eff2}. 

We first consider a Hermitian case with two different dielectric (ceramic) cylinders~\cite{note1,ceramic} where $\varepsilon'_L=\varepsilon'_R/1.44$ and $\varepsilon''_L=\varepsilon''_R=0$. The CBS calculated from the dispersion relation [Eq.~(\ref{eqn:DRC})] between the real frequency and the complex normalized wavevector $K=k_ya/(2\pi)$ is plotted in Fig.~\ref{scm}(b). The broken symmetries lead to asymmetric $\text{Re}(K)$~(black lines), while $\text{Im}(K)$~(blue lines) is symmetric due to the lossless condition. This is because the two complex solutions of $K$ at fixed $\omega$ must be either purely real or form a pair of complex conjugates~[i.e., $\omega(K)=\omega({K}^*)$] when the system is Hermitian~\cite{Supp}. Such spectral nonreciprocity is most profound in the range between 3.6~GHz and 4~GHz, which is due to the two nonreciprocal local resonance frequencies of single-grating layers. In comparison, symmetric dispersion appears near the Bragg gap at frequency $f>4.6$~GHz. It is apparent that non-zero imaginary $K$ appears in the region of bandgaps.

\begin{figure}[!tbp]
	\includegraphics[width=3.5in]{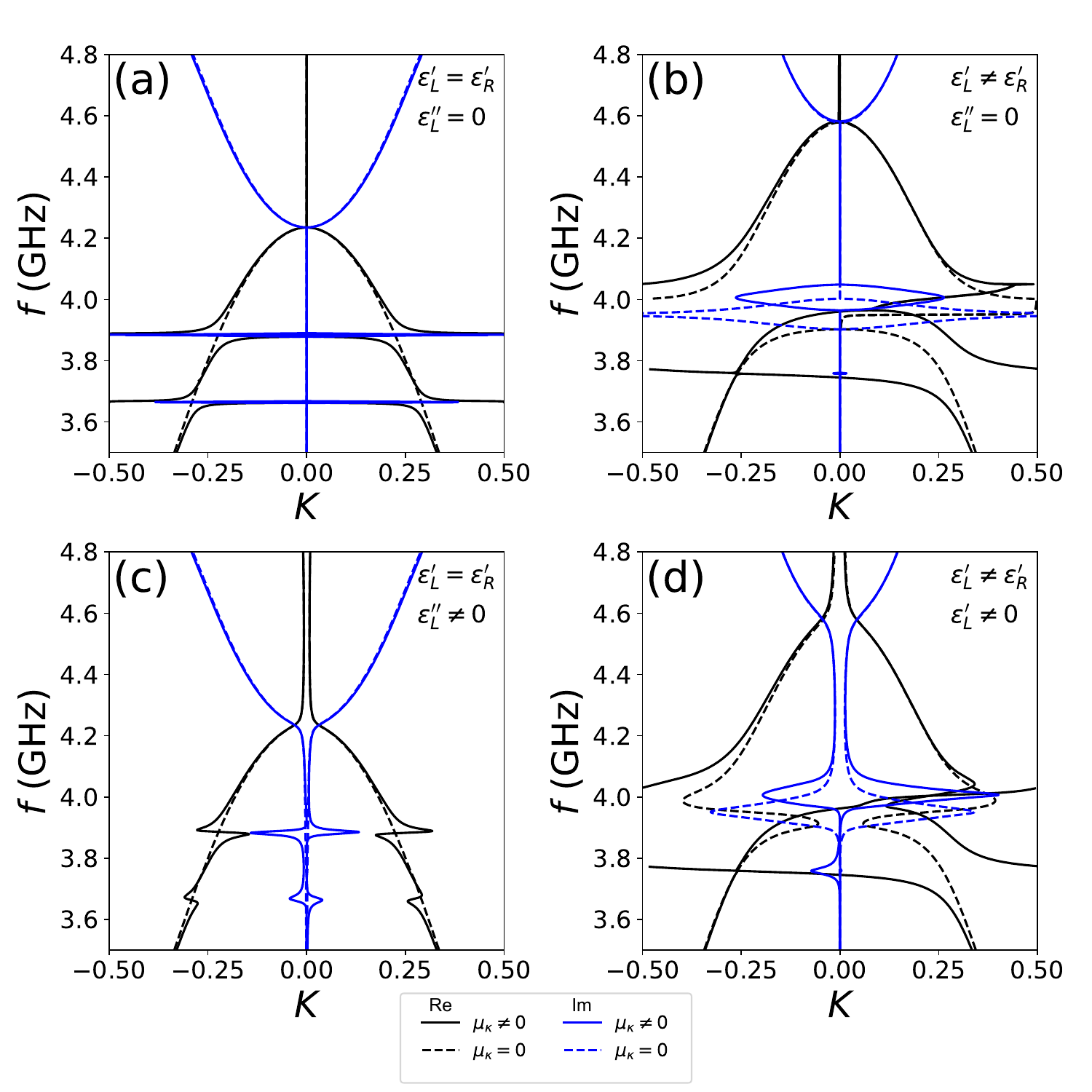}
	\centering
	\caption {\small (color online) Effects of spatial asymmetry, non-Hermiticity, and nonreciprocity on complex band structures. (a) Trivial Hermitian case. (b) Nontrivial Hermitian case. (c) Trivial non-Hermitian case. (d) Nontrivial non-Hermitian case. Black and blue lines correspond to, respectively, the real and imaginary parts of $K$. For direct comparison, results for reciprocal ($\mu_\kappa=0$) and non-reciproal ($\mu_\kappa\neq0$) cases are shown as dashed lines and solid lines, respectively. The parameters are $\varepsilon_L'=\varepsilon_R'/1.44$ for nontrivial cases and $\varepsilon_L''=2$ for non-Hermitian cases. Other parameters are the same as in Fig.~\ref{scm}.
}
	\label{fig:cbs}
\end{figure}

To understand the band hybridization near local resonance frequencies, we compare four scenarios involving Lorentz nonreciprocity, which are summarized in Table~\ref{tb1} with the corresponding CBSs being shown as solid lines in Fig.~\ref{fig:cbs}. We note that the ``trivialness" defined in Table~\ref{tb1} refers only to symmetry instead of topology. Lorentz reciprocity (i.e., the condition of $\boldsymbol{\mu}=\boldsymbol{\mu}^\text{T}$) can further be ``switched" on or off by changing the parameter $\mu_\kappa$. For a complete comparison, the results for the corresponding Lorentz reciprocal~($\mu_\kappa=0$) cases are also shown as dashed lines for each of the four cases mentioned above. It should be noted that the nontrivial Hermitian case~[Fig.~\ref{fig:cbs}(b)] is identical to that in Fig.~\ref{scm}(b) except for the addition of dashed lines. 

\begin{table}[]
\centering

\caption{Classification of the four general cases with $\mu_\kappa\neq0$.}
\label{tb1}
\begin{tabular}{c |c c}
\hline \hline
Dielectric & Trivial & Non-trivial \\
parameters & $\varepsilon_L'=\varepsilon_R'$   & $\varepsilon_L'\neq\varepsilon_R'$ \\
\hline
Hermitian $\varepsilon_L''=0$  & Case (a) & Case (b) \\
Non-Hermitian $\varepsilon_L''\neq0$ & Case (c) & Case (d) \\
\hline\hline
\end{tabular}

\end{table}

In the trivial Hermitian case [Fig.~\ref{fig:cbs}(a)], the imaginary part of CBS~(blue solid lines) shows two sharp peaks of $|\text{Im}(K)|$ at around 3.7~GHz and 3.9~GHz, which are associated with local resonances. Narrow gaps associated with divergingly large $\text{Im}(K)$ are opened due to the hybridization among propagating photonic modes and the local resonances. These local resonances are nonreciprocal due to the breaking of Lorentz reciprocity by magnetized materials. However, the band structure is still symmetric [i.e. $\omega(K)=\omega(-K)$]. If we replace the nonreciprocal relative permeability tensor with a symmetric one by setting $\mu_\kappa=0$, these local resonances gap disappear from the frequency range in Fig.~\ref{fig:cbs}(a) (dashed lines), showing only one single continuous band below 4.2~GHz. In contrast, no noticeable change is observed for the trivial Bragg gap above 4.2~GHz.

In the nontrivial Hermitian case, the left-right symmetry is broken by a reduction in the dielectric constant of the left cylinder (i.e., $\varepsilon_L'=\varepsilon_R'/1.44$). The CBS is shown as solid lines in Fig.~\ref{fig:cbs}(b), which is a replication of the same plot in Fig.~\ref{scm}(b). In addition to the lifting of bands to higher frequencies due to lower index, this also leads to strong asymmetry in the real part of CBS (as mentioned previously) and broadening of the local resonance bandgap near 4~GHz. For comparison, we include the dashed lines in Fig.~\ref{fig:cbs}(b) to represent a fully symmetric CBS of the ``non-magnetized" case by setting $\mu_\kappa=0$. It is shown that the real part of CBS in the magnetized case~(black solid lines) is generally symmetric except for the range of strong band hybridization. As mentioned above, the imaginary part of CBS~(blue solid lines) shows symmetric peaks in $\text{Im}(K)$ even though the real part is not symmetric. In contrary to the common belief on the strong relation between nonreciprocal transport and band properties~\cite{spec_non,spec_non2}, we find that such asymmetry in $\text{Re}(K)$ will not contribute to the transmittance, which is discussed in a later section of this Letter.

\begin{figure}[!tbp]
	\includegraphics[width=3.3in]{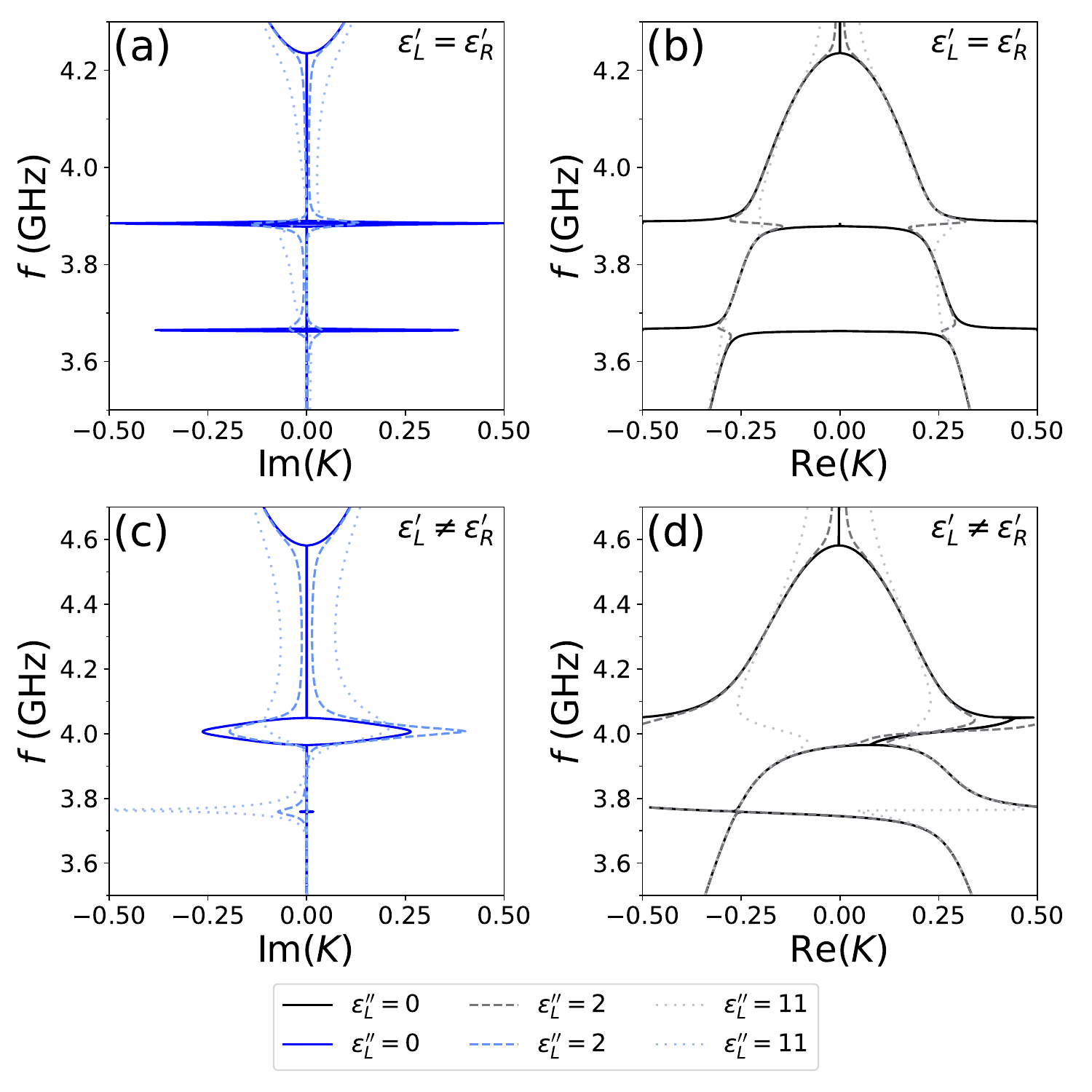}
	\centering
	\caption {\small (color online) Effects of non-Hermiticity on nonreciprocal bands for trivial and nontrivial cases. (a)~Imaginary part and (b)~Real part of CBS where $\varepsilon_L'=\varepsilon_R'$ (trivial nonreciprocal case). (c)~Imaginary part and (d)~Real part of CBS where $\varepsilon_L'=\varepsilon_R'/1.44$ (nontrivial nonreciprocal case). The solid lines, dashed lines, and dotted lines refer to $\varepsilon_L''=$ 0, 2, and 11, respectively. Other parameters are the same as in Fig.~\ref{scm}.
}
	\label{fig:loss}
\end{figure}

To study the effect of non-Hermiticity, we introduce energy loss~($\varepsilon_L''=2$) to the Hermitian photonic crystals already discussed in Figs.~\ref{fig:cbs}(a) and~\ref{fig:cbs}(b). The corresponding CBSs with such non-Hermitian effect are shown in Figs.~\ref{fig:cbs}(c) and~\ref{fig:cbs}(d). In both trivial (c) and nontrivial (d) cases, the results clearly show asymmetries in both $\text{Re}(K)$ and $\text{Im}(K)$. Since conjugate pairing of $K$ is no longer required in non-Hermitian cases, the diverging frequencies in $\text{Im}(K)$ for the two directions become misaligned, and the waves decay differently in the two opposite wave propagation directions. To further illustrate the effect, we separate the real and imaginary parts of Figs.~\ref{fig:cbs}(b) and~\ref{fig:cbs}(d) and show their changes with respect to $\varepsilon_L''$ in Fig.~\ref{fig:loss}. The new curves for $\varepsilon_L''=11$ show that extreme $\varepsilon_L''$ can lead to highly asymmetric CBS with asymmetric peaks of $\text{Im}(K)$. However, there is an obvious difference between the trivial and non-trivial cases. In the trivial case [Figs.~\ref{fig:loss}(a),(b)], we can see weakening in the resonance strength and broadening of the resonance linewidth. When the absorption parameter increases to $\varepsilon_L''=11$, there are much clearer asymmetries in the band regions compared to the gap regions. In contrast, the non-trivial case [Figs.~\ref{fig:loss}(c),(d)] shows obvious further diverging $\text{Im}(K)$ near the mini gap at 3.75 GHz. In both cases, there is a large distortion in the real parts of the CBS on one side when $\varepsilon_L''$ is large, and we find no obvious relation between $\text{Re}(K)$ and the magnitude of $\text{Im}(K)$.

\begin{figure}[!tbp]
	\includegraphics[width=3.3in]{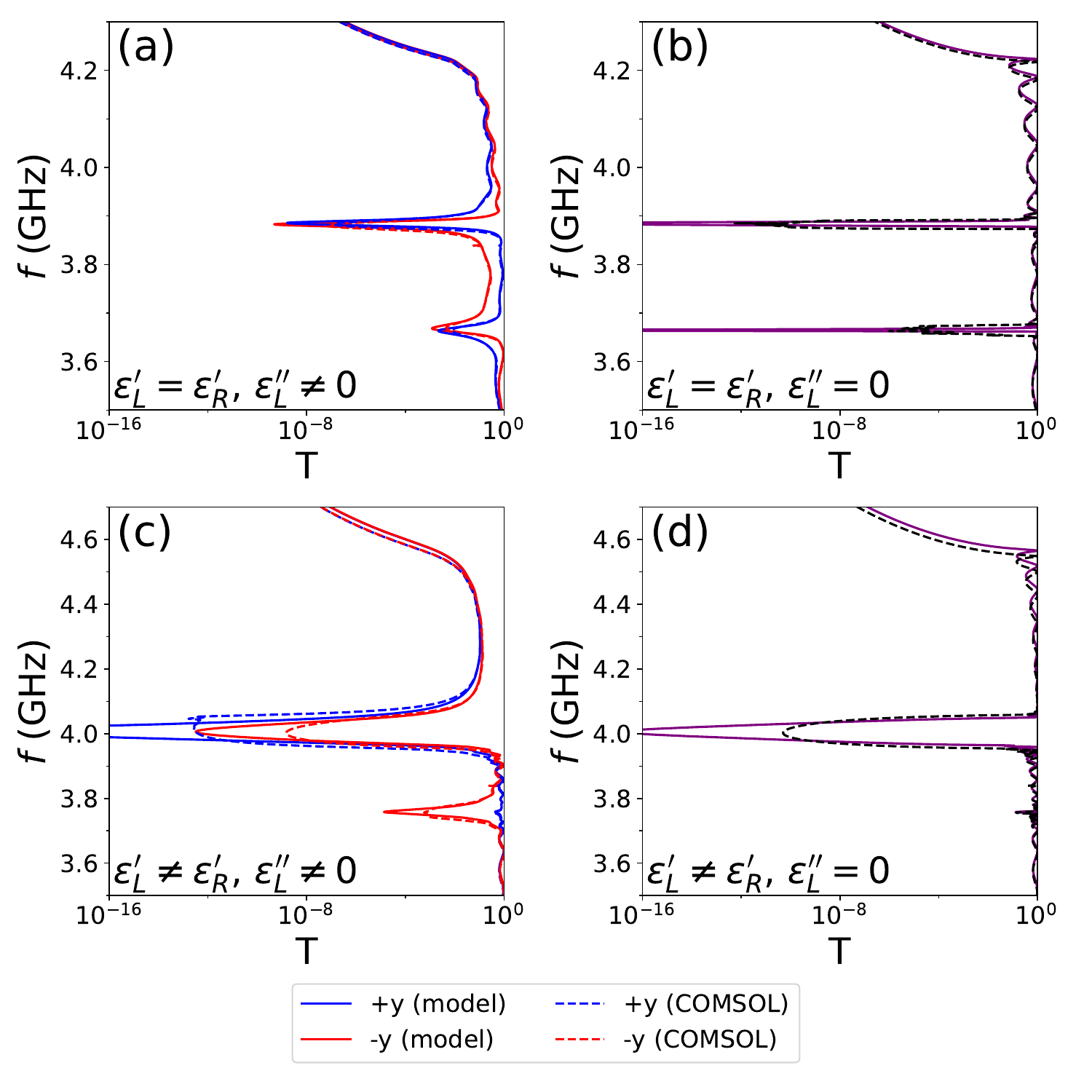}
	\centering
	\caption {\small (color online)~Effects of non-Hermiticity on transmittance spectra. (a) Trivial non-Hermitian case with $\varepsilon_L'=\varepsilon_R'$ and $\varepsilon_L''= 2$. (c) Nontrivial non-Hermitian case with $\varepsilon_L'=\varepsilon_R'/1.44$ and $\varepsilon_L''= 2$. Number of layers $N=12$. Other parameters are the same as in Fig.~\ref{scm}. The solid lines and dashed lines indicate the transmittance calculated from our analytical model and the full-wave simulation in COMSOL MULTIPHYSICS, respectively. The color in red and blue denote, respectively, the results for $-y$ (backward) and $+y$ (forward) propagation directions. The corresponding Hermitian cases with $\varepsilon_L''= 0$ for (a) and (c) are shown in (b) and (d), respectively. The solid purple lines indicate two identical overlapping analytical results from Eq.~(\ref{eqn:trm}) for both forward and backward incident directions while the dashed black lines correspond to that of the full-wave simulation results.
}
	\label{fig:Tml}
\end{figure}

To reveal the meaning of the complex wavenumber in transmission, we obtain a closed form expression of the transmittance $T$ based on Eq.~(\ref{eqn:Cd}) as~\cite{Supp}
\begin{equation}
    T_\pm={|t_N^\pm|}^2=\left|\frac{4z}{(1+z)^2e^{\pm ik_y^\pm Na}-(1-z)^2e^{\pm ik_y^\mp Na}}\right|^2,\label{eqn:trm}
\end{equation}
where
\begin{equation}
    z=\sqrt{\frac{(1+\rho e^{ik_0a})^2-\tau^+\tau^-e^{i2k_0a}}{(1-\rho e^{ik_0a})^2-\tau^+\tau^-e^{i2k_0a}}}
\end{equation}
has the meaning of the relative impedance of the system in effective medium description~\cite{nm}, $N$ is the number of layers, and $T_\pm$ and ${t_N}^\pm$ denote the transmittance and the transmission coefficient for the forward and backward solutions. It can be shown from Eq.~(\ref{eqn:trm}) that the isolation ratio:
\begin{equation}
    \left|\text{ln}\left(\frac{T_+}{T_-}\right)\right|=4\pi N  \left|\Delta\text{Im}(K)\right|,\label{eqn:tratio}
\end{equation}
is fully determined by the difference in the magnitudes of imaginary parts of wavenumbers $\Delta\text{Im}(K)\equiv|\text{Im}(k_y^+a/2\pi)|-|\text{Im}(k_y^-a/2\pi)|$. Here, we have used the fact that the pair of forward and backward solutions in Eq.~(\ref{eqn:DRC}) should have opposite signs in $\text{Im}(k_y)$.

Finally, we show the transmittance spectra calculated using Eq.~(\ref{eqn:trm}) for the magnetized cases~(see solid lines in Fig.~\ref{fig:Tml}). The results align closely with the full-wave finite-element simulation shown as dashed lines, except that there are small differences due to the non-zero mesh size of the FEM and the dipole approximation in our analytical model. The results indicate that the forward transmittance (blue line) differs from the backward one~(red line) in both trivial~[Fig.~\ref{fig:Tml}(a)] and non-trivial~[Fig.~\ref{fig:Tml}(c)] cases when the system is non-Hermitian (i.e., $\varepsilon_L''\neq 0$). The numerical results are consistent with our analytical formula in Eq.~(\ref{eqn:tratio}) and the two curves of $\text{Im}(K)$ shown as blue dashed lines in Figs.~\ref{fig:loss}(a) and ~\ref{fig:loss}(c). For the trivial non-Hermitian case in Fig.~\ref{fig:Tml}(a), an interesting nonreciprocal transmittance window appears between 3.65~GHz and 3.9~GHz with forward transmittance greater than backward transmittance. Such an asymmetric attenuation of waves can also be understood as selective absorption due to the difference in field strength between the two dielectric cylinders~\cite{tms2}. For the nontrivial non-Hermitian case in Fig.~\ref{fig:Tml}(c), we see a backward transmittance dip at around 3.8~GHz, which align with the corresponding diverging negative $\text{Im}(K)$ in Fig.~\ref{fig:loss}(c), while the forward transmittance maintains a relatively stable value due to small positive $\text{Im}(K)$. Both our analytical model and full-wave simulation confirm the absence of nonreciprocal transmittance when the system is Hermitian, as shown in Figs.~\ref{fig:Tml}(b) and~\ref{fig:Tml}(d) for $\varepsilon_L''= 0$.

In conclusion, we have studied the band hybridization associated with nonreciprocal local resonances in magnetophotonic crystals using a complex band structure perspective. Nonreciprocal local resonances combined with sufficient symmetry breaking could selectively provide a divergingly large $\text{Im}(\mathbf{k})$ in the backward transport of light. Our analytical solution for a 2D photonic system reveals a commonly misunderstood aspect in light transport in nonreciprocal photonic systems. These analytic results are consistent with full-wave simulations. Our example demonstrates that the commonly believed statement that ``spectral nonreciprocity [$\omega(\mathbf{k})\neq\omega(-\mathbf{k})$] implies nonreciprocal transmission" is not always correct. Furthermore, we have shown that the isolation ratio for nonreciprocal transmission is determined solely by $\text{Im}(\mathbf{k})$, instead of the extensively studied real part $\text{Re}(\mathbf{k})$. We note that, in general, the signal transport speed of a non-Hermitian system cannot be inferred from the slope ($\partial\omega/\partial k$) in the CBS without generalizing it to complex group velocity~\cite{Wolff}.  We conclude that non-Hermiticity plays a crucial role in controlling the transmission nonreciprocity, which can only be revealed by examining the complex band structures with complex values of $\mathbf{k}$. 

We thank Yong-Liang Zhang, Jin Wang, Xiaohan Cui, Ruo-Yang Zhang, and Zhaoqing~Zhang for their assistance and fruitful discussions. This work was supported by the Hong Kong Research Grants Council through project nos. AoE/P-502/20 and C6013-18G.

\bibliographystyle{apsrev4-2}
\bibliography{refnew}

\begin{thebibliography}{46}
\bibitem [{not()}]{note1}%
  \BibitemOpen
  \bibinfo {note} {The parameters are similar to Ref.~\cite{tms2}, with
  dielectric ceramics in the Appendix 2 of Ref.~\cite{ceramic}}\BibitemShut
  {NoStop}%
\bibitem [{Com()}]{Comment1}%
  \BibitemOpen
  \bibinfo {note} {Evanescent coupling between gratings is neglected in
  obtaining the closed-form solution since the distance between gratings is
  almost five times the maximum decay length of the evanescent modes for the
  frequency range. If the distance between gratings is smaller instead, nonzero
  evanescent coupling between gratings cannot be neglected.}\BibitemShut
  {Stop}%
\end{thebibliography}
\end{document}